\documentclass[iop]{emulateapj}                             
\newcommand{\figscaleone}{\epsscale{1.15}}      
\newcommand{\figscaletwo}{\epsscale{0.72}}       

\newcommand{\htwo}{${\rm H_2}$}
\newcommand{\hsix}{${\rm H_6^+}$}
\newcommand{\hd}{${\rm (HD)_3^+}$}

\newcommand{\hminus}{${\rm H^-}$}
\newcommand{\percm}{${\rm\,cm^{-1}}$}
\newcommand{\mum}{$\,{\rm\mu m}$}

\newcommand{\B}[1]{\ensuremath{\mathbf{#1}}}
\newcommand{\R}[1]{\ensuremath{\mathrm{#1}}}

\slugcomment{Submitted to ApJ 21st February 2011; Revised 4th April 2011 and 6th May 2011}

\shorttitle{Interstellar solid hydrogen}
\shortauthors{Lin {\it et al\/}}

\begin{document}

\title{Interstellar solid hydrogen}

\author{Ching Yeh Lin, Andrew T.B. Gilbert}
\affil{Research School of Chemistry, Australian National University, Canberra ACT 0200, Australia}
\email{cylin@rsc.anu.edu.au,andrew.gilbert@anu.edu.au}

\author{Mark A. Walker}
\affil{Manly Astrophysics, 3/22 Cliff Street, Manly 2095, Australia}
\email{Mark.Walker@manlyastrophysics.org}

\begin{abstract}
We consider the possibility that solid molecular hydrogen is present in interstellar space. If so cosmic-rays and energetic photons cause ionisation in the solid leading to the formation of \hsix. This ion is not produced by gas-phase reactions and its radiative transitions therefore provide a signature of solid \htwo\  in the astrophysical context. The vibrational transitions of \hsix\  are yet to be observed in the laboratory, but we have characterised them in a quantum-theoretical treatment of the molecule; our calculations include anharmonic corrections, which are large. Here we report on those calculations and compare our results with astronomical data. In addition to the \hsix\ isotopomer, we focus on the deuterated species \hd\  which is expected to dominate at low ionisation rates as a result of isotopic condensation reactions. We can reliably predict the frequencies of the fundamental bands for five modes of vibration. For \hd\ all of these are found to lie close to some of the strongest of the pervasive mid-infrared astronomical emission bands,  making it difficult to exclude hydrogen precipitates on observational grounds. By the same token these results suggest that \hd\ could be the carrier of the observed bands. We consider this possibility within the broader picture of ISM photo-processes and we conclude that solid hydrogen may indeed be abundant in astrophysical environments.  
\end{abstract}


\section{Introduction}
Astronomers observe matter in a great variety of physical conditions and, correspondingly, hydrogen exists in a variety of states in different astronomical contexts. Liquid metallic hydrogen is inferred to be present in a degenerate core in the planets Jupiter and Saturn \citep{hubbard}, and in the interstellar context there are known to be three phases of hydrogen in rough pressure equilibrium: hot, ionised, low-density plasma; atomic gas; and molecular hydrogen in cold, dense, interstellar clouds. Might there also be solid molecular hydrogen? 

That question was first addressed in the late 1960's when it was suggested that thick mantles of solid \htwo\ might be deposited onto the surfaces of graphite grains --- see, for example, \citet{wickramasinghe1,hoyle, wickramasinghe2}. But these early papers employed a poor description of the \htwo\ saturated vapour pressure and in fact solid \htwo\ mantle formation is not expected under the conditions known to exist in the diffuse ISM \citep{greenberg,field}. The difficulty is simply that the pressure in the diffuse interstellar medium is much smaller than the saturated vapour pressure of \htwo, even for temperatures as low as  that of the microwave background.

The argument just given may, however, be misleading because the phase equilibrium of \htwo\ in astrophysical environments may differ significantly from that of pure \htwo. In particular the electrostatic binding energy contributed by ions in the lattice may substantially increase the total binding energy of the solid, even at very low ion concentrations. Astrophysical \htwo-ice may thus be more robust than is commonly supposed.

It also remains a possibility that there exists gas which is so cold and dense that pure hydrogen precipitates can form. But that gas cannot be part of the diffuse ISM and must, instead, form distinct, self-gravitating entities. Clouds of this type -- cold, dense, mostly transparent and having very little thermal emission -- were postulated by Pfenniger and Combes (1994), Pfenniger, Combes and Martinet (1994), to be a significant component of the Galaxy's dark matter. That picture was subsequently considered by many other authors (e.g. Gerhard and Silk 1996, Walker and Wardle 1998, Draine 1998, Sciama 2000, Walker 2007). It was argued by \citet{wardle} that small amounts of solid \htwo\ can confer thermal stability on these clouds. 

Irrespective of these theoretical deliberations, observation provides the final arbiter of whether solid \htwo\ is present in the Galaxy. To that end we must first ask how solid \htwo\ might be detected if it were present? That is the question addressed in this paper. It is not a trivial matter because solid \htwo, like the gas from which it forms, is cold, dense and very nearly transparent in the optical. We make no attempt at a comprehensive assessment of the detection problem; instead we focus on the properties of a particular molecule, \hsix, which is known to form in solid \htwo\ exposed to ionising radiation (see \S2). Cosmic-rays and energetic photons pervade  interstellar space and we therefore expect \hsix\ to be present at some level in solid, interstellar \htwo. Furthermore, this molecular ion does not form in any significant quantities in the gas-phase chemistry of \htwo, so it is a good tracer of solid \htwo.

Currently \hsix\ is not well characterised experimentally so in order to predict its astrophysical manifestations we are first obliged to confront problems in  quantum chemistry. To tackle these we have utilised the Q-Chem\footnote{http://q-chem.com} software package (Shao et al. 2006), with some bespoke extensions, for our calculations. For the most part we have concentrated on the bare \hsix\ molecule, whereas in solid \htwo\ it is expected that the ion forms many weak bonds with surrounding \htwo\ molecules, leading to a much larger complex. Our focus on bare \hsix\ is motivated by the fact that solvation does not greatly disturb the structure of the \hsix\  -- see \citet{kurosaki2} -- and thus the high frequency vibrational transitions of the solvated forms should closely resemble those of the bare molecule. Moreover we do not yet know exactly what form the solvation takes. We do, however, give some consideration to how the properties of the solvated forms differ from those of the bare molecule, and where possible we estimate the size of these effects. This has been done by reference to the properties of ${\rm H_{14}^+=H_6^+(H_2)_4}$ --- a cluster which is much smaller than those which are likely to occur in practice.

We have neglected any other interactions between \hsix\ and the host lattice. This is a reasonable assumption for an initial exploration as the intermolecular forces in solid hydrogen are very feeble. Indeed solid {\it para\/}-\htwo\ is commonly used for matrix isolation spectroscopy of small molecules (see, e.g., Oka 1993, Anderson et al 2002), and this is precisely because it permits nearly free vibration and rotation of the impurity species. In the astronomical context there is plenty of time for {\it ortho-para\/} conversion to take place, so any cold \htwo\ is expected to be predominantly in the rotational ground state and is thus in the {\it``para''\/} form.

This paper is structured as follows. In the next section we discuss relevant laboratory results on irradiated solid hydrogens and their interpretation. In \S3 we present our theoretical model of the \hsix\ molecule and its isotopic variant \hd. The latter is expected to be the most abundant isotopomer at low ionisation rates whereas the former dominates at high ionisation rates. Section 4 compares the computed vibrational transitions of \hsix\  and \hd\ with mid-IR astronomical observations. The properties of ten other isotopomers, which may be of interest in laboratory studies, are tabulated in Appendix A.  Similarities between computed and observed spectra prompt us to suggest that \hd\ may actually be responsible for the observed mid-IR astronomical emission bands; the implications of this suggestion are discussed in \S5 and we offer our conclusions in \S6.

\section{Lab studies of H$_6^+$ in irradiated, solid \htwo}
At temperatures less than the triple-point (13.8~K), hydrogen is able to form a low-density solid ($\rho\simeq0.086\;{\rm g\,cm^{-3}}$). It is a ``quantum solid'', meaning that the zero-point motion of the molecules is a large fraction of the intermolecular separation. Solid hydrogens have been extensively studied in the laboratory; see \cite{silvera}  and \cite{souersbook} for reviews of their properties; \citet{vankranendonk} provides theoretical models for some of these properties.

Although the natural abundance of tritium is negligible, its importance in fusion research led to studies of solid hydrogens in which tritium is incorporated. These solids naturally exhibit the effects of irradiation because tritium undergoes beta decay to ${\rm^3He}$ with a half-life of approximately 12 years. The energetic electrons which result from beta decay cause ionisation and dissociation of the \htwo. Various radiation-induced phenomena have received attention in the literature, as described below.

\subsection{Radiation-induced absorption lines}
In the first studies of irradiated solid hydrogens it was discovered that some extra lines were present in the absorption spectrum near the fundamental vibrational band of the solid \citep{souers1,souers2}. These lines were attributed to the presence of ions in the matrix which induce a Stark-shift in the quantum states of neighbouring molecules.

Detailed studies of these spectral features were made by \citet{brooks1,brooks2} using proton-beam irradiation of tritium-free solid ${\rm D_2}$ --- an approach which allows complete control over the magnitude and timing of the radiation doses. The radiation-induced lines were seen to persist for many hours after irradiation ceased, implying that the ions involved have very low mobilities. \cite{brooks2} identified the immobile anion as an ``electron bubble'' (i.e. an electron trapped in a lattice defect)  but did not claim to have identified the immobile cation; they suggested ${\rm D_3^+}$, or the solvated form ${\rm D_9^+=D_3^+(D_2)_3}$.

Later studies of electron-beam and gamma-ray irradiated solids by Oka and collaborators \citep{oka,chan,momose} emphasised the use of {\it para}-\htwo\ (i.e. $J=0$) enriched solids to achieve intrinsically very narrow spectral lines; again the cations were interpreted as ${\rm H_3^+}$ and its solvated forms.

\subsection{Broad spectral features}
In parallel with the investigations of radiation-induced absorption lines there were studies of broad-band absorption in irradiated solid hydrogens. Again these studies commenced with solids containing tritium \citep{richardson, poll}, and later utilised external sources of ionising radiation \citep{selen, forrest1}. Broad infrared/optical absorptions were reported and interpreted in terms of bound-bound transitions of electron bubbles  \citep{richardson,poll,selen}. And a continuum UV absorption was reported and interpreted as the bound-free transitions of electron bubbles \citep{forrest1}.

Surprisingly, irradiated solid hydrogens were also observed to {\it emit\/} optical/near-infrared continua in a broad band peaking around 825$\,$nm \citep{forrest2,magnotta,forrest3}. Both steady emission and short flashes were seen to occur, with essentially the same spectrum. Emission from solid deuterium was observed to be much stronger than that from solid \htwo.  It was also observed that the optical emission increased in response to UV exposure, indicating a close connection with the UV continuum absorption process \citep{magnotta}. But the actual emission process was not securely identified. One possibility is that the emission is due to radiative attachment of ${\rm e^-}$ and hydrogen atoms \citep{forrest2,forrest3}. In gas-phase the \hminus\ bound-free opacity (e.g. Doughty, Fraser and McEachran 1966) has a spectral peak in a similar location to the observed emission from solid D$_2$, but the ${\rm H^-}$ opacity peak is much broader than that of the observed emission. 

\subsection{Paramagnetic signals}
By virtue of having no unpaired electrons \htwo\ molecules themselves have no Electron Spin Resonance (ESR) signal. But the irradiated solid exhibits several features (e.g. Kumada et al 2008). First there is a strong signal from atomic hydrogen, produced by dissociation of the molecules. Secondly there is a signal from free electrons in the solid, produced by ionisation of the molecules. And thirdly there is a quartet of lines \citep{miyazaki}, initially suggested to be due to ${\rm H_2^-}$ \citep{symons1}, but later recognised as the signature of \hsix\ \citep{symons2,suter,kumada1}. Further ESR investigations of irradiated solid hydrogens conclusively demonstrated that \hsix\ is indeed the source of the quartet lines by studying solid mixtures of \htwo\ and HD or ${\rm D_2}$, for which a variety of isotopomers of \hsix\ arise, each with its own unique ESR signature \citep{kumada2,kumagai}.

\subsection{Molecular hydrogen ion clusters}
All of the literature cited in \S\S2.1,2.2 implicitly assumes that molecular hydrogen ion chemistry in the condensed phase resembles that of the gas phase, for which it is well established that clusters ${\rm H}_n^+$ with odd-$n$ occur in much greater abundance than those with even-$n$ \citep{clampitt,hiraoka,ekinci}. The subsequent discovery of \hsix\ in the ESR experiments on irradiated solid hydrogens (\S2.3) challenges that assumption. There seems to be no experimental data which point specifically to the presence of ${\rm H_3^+}$, or its solvated forms, in irradiated solid hydrogens: it is possible that the dominant cations are \hsix\ isotopomers and their solvated forms.

The existence of even-$n$ clusters was predicted on the basis of {\it ab initio\/} calculations which showed them to have comparable binding energies to the odd-$n$ clusters \citep{wright}. The even-$n$ clusters were then discovered in gas phase via mass spectroscopy \citep{kirchner}, albeit at abundances very much lower than those of odd-$n$. \hsix\ was seen in these experiments to have much higher abundance than ${\rm H_4^+,\,H_8^+\; or \;H_{10}^+}$  and this result prompted further theoretical investigations of \hsix\ and the other even-$n$ clusters \citep{montgomery,kurosaki1,kurosaki2}. In particular \citet{kurosaki1} confirmed the stability of \hsix\ and identified its most stable configuration as $D_{2d}$ symmetry --- a structure which is {\it not\/} based on the ${\rm H_3^+}$ moiety, as can be seen in figure 1.

Na\"ively \hsix\  can be thought of as a central H$_2^+$ with an \htwo\ molecule bonded perpendicularly at each end, the axes of the three moieties being mutually orthogonal. In fact each atom in the molecule carries a similar fraction of the total charge \citep{kurosaki2}. These authors also demonstrated the stability of larger clusters, up to ${\rm H_{14}^+}$. The structure of the clusters was shown to correspond to solvation of \hsix\ by \htwo, with the solvating molecules being loosely bound and carrying very little charge, and the central \hsix\ moiety only slightly perturbed in its charge distribution and bond lengths. 

The existence of very large even-$n$ clusters has been demonstrated experimentally by \citet{jaksch}: molecular hydrogen ion clusters as large as $n=120$ were observed via mass spectroscopy of ionic species formed inside ultra-cold (0.4~K) helium nanodroplets. Odd-$n$ species were also produced and the ratio of even:odd clusters was observed to be small. \citet{jaksch} argued that their high yield of even-$n$ clusters indicated that cluster ionisation dynamics are affected by a helium matrix. The ionisation dynamics of hydrogen clusters have been studied theoretically by \citet{tachikawa}.

\begin{figure}
\figscaleone
\plotone{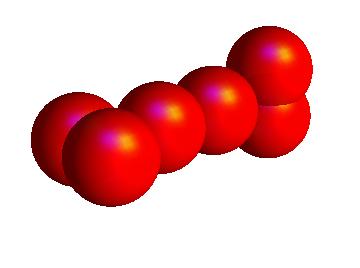}
\caption{The $D_{2d}$ configuration of \hsix, which is the structure corresponding to the global minimum of potential energy for this molecule \citep{kurosaki2}.}
\end{figure}

\subsection{Summary of impurities in solid \htwo}
The ESR studies described in \S2.3 demonstrate explicitly that irradiated solid hydrogens contain free electrons, atomic hydrogen and \hsix. Free electrons and H-atoms can combine to yield \hminus, and this reaction is strongly favoured by the large ratio of the electron affinity of hydrogen (755~meV) to the temperature ($T<14\,$K, so $kT\la1\,$meV). In a low temperature bath of \htwo, solvation of this anion and the \hsix\ cation are both expected. Icosahedral solvation shells have been suggested for both species \citep{wang,jaksch}. But it is not clear whether this remains a preferred geometry for solvation in the context of a solid \htwo\ matrix, where the interface between the solvation shells and the \htwo\ lattice presumably plays a significant role in determining the form of solvation. 

In the astrophysical context hydrogen is not expected to occur as a pure element, but mixed with helium and trace amounts of metals. It appears that helium does not alloy with solid hydrogen \citep{leventhal}, but it is possible that a solid hydrogen matrix might contain metal impurities. In this paper we ignore that possibility and confine our attention to the hydrogen alone as there is plenty to occupy us even in this restricted case. We further assume that the deuterium abundance is $X_{\rm D}\sim10^{-5}$ by number, i.e. comparable to the deuterium abundances measured in diffuse Galactic gas (e.g. Sembach 2010). We expect this to manifest itself as an abundance of ${\rm HD:H_2}$ that is $X_{\rm D}$ by number.

A solid hydrogen of this composition, when subject to ionising radiation, will usually yield \hsix\ that is free of deuterium at the time of formation. However, the Zero-Point Vibration Energy (ZPVE) of singly-deuterated \hsix\ (plus \htwo) is about 13~meV lower than that of the undeuterated isotopomer (plus HD) -- see Appendix A -- so at low temperatures condensation of HD onto the \hsix\ proceeds. And the net ZPVE is further lowered by a similar amount for each of two subsequent deuterations, with the result that at temperatures less than about 12~K condensation proceeds all the way to \hd.

We use the shorthand \hd\ to denote the specific isotopomer ${\rm[HD(HD)HD]^+}$, where the central- and two end-moieties are encoded in the notation as ${\rm[End(Centre)End]^+}$.

If the ionisation rate in the solid is sufficiently high that the concentration of \hsix\ is of order $X_{\rm D}/3\sim3\times10^{-6}$, then there is insufficient HD in the lattice to fully deuterate the \hsix\ and at these concentrations the singly- and doubly-deuterated isotopomers listed in Appendix A will constitute a significant fraction of the \hsix. At still higher ionisation rates even these lightly-deuterated isotopomers are in the minority and the \hsix\ isotopomer dominates. For simplicity we consider here only very high and very low ionisation rates, for which the \hsix\ is essentially all in the form of the \hsix\ and \hd\ isotopomers, respectively. 

The first two HD condensations on \hsix\ were observed by \citet{kumagai} in ESR studies of irradiated solid hydrogens at 4~K, but these authors did not report a detection of \hd. As each of the deuteration reactions involves lowering the ZPVE by a similar amount there is an expectation that if condensation starts it will go all the way to \hd. We therefore assume that \hd\ was not detected by \citet{kumagai} because its ESR spectrum is very complex and/or the lines are broad. We note that \citet{kumagai} did not report any isotopomer of \hsix\ containing ${\rm D_2}$ or ${\rm D_2^+}$ moieties except when the solid itself was made from a mixture including ${\rm D_2}$. This demonstrates that the condensation process is one where diatomic hydrogen molecules are preserved as units which are substituted in or out of the \hsix\ --- a situation which is unsurprising considering the large energy (4.5~eV) that is required to break up \htwo. 

\section{Theoretical model of \hsix}
Before tackling \hsix\ itself we performed preliminary theoretical studies on H$_2$ and H$_2^+$. We used various {\it ab initio\/} electronic structure methods, all available within the Q-Chem software package \citep{shao}, to calculate the bond length and fundamental harmonic frequency for each molecule.  From these calculations it was determined that the most suitable level of theory for our calculations is CCSD (Coupled Cluster with Single and Double excitations), combined with  cc-pVTZ basis functions (correlation-consistent polarised Valence-only Triple Zeta-function). We found that the CCSD/cc-pVTZ combination gave a good compromise between accuracy and computational cost, for both H$_2$ and H$_2^+$, with deviations of less than 0.001\AA{} and 5\percm\ from our most accurate method, CCSD/cc-pVQZ, which utilises a Quadruple Zeta-function basis set.  Henceforth all the results we quote in this paper were obtained by using the CCSD/cc-pVTZ level of theory, unless stated otherwise.

To determine the importance of augmenting our basis function set with additional diffuse functions, we performed calculations of the energy required to break \hsix\ into 2H$_2+$H$_2^+$. The binding energy at the CCSD/cc-pVTZ level is $235.57\;{\rm kJ\,mol^{-1}}$ whereas the CCSD/aug-cc-pVTZ energy is $236.35\;{\rm kJ\,mol^{-1}}$.  The effect of augmenting our chosen basis is thus one third of one percent, from which we conclude that diffuse functions are not required.

After our calculations on \hsix\  were completed we became aware of a comprehensive {\it ab initio\/} study of the properties of \hsix\  \citep{hao}, which had been published shortly after we completed our initial survey of the literature. The study of \citet{hao} took in six different possible structures for \hsix\ and utilised various levels of theory: both Restricted and Unrestricted Hartree-Fock theory, and CCSD with extension to perturbative and full triple-excitations.  They used the basis sets cc-pVTZ and cc-pVQZ and their augmented counterparts aug-cc-pVTZ and aug-cc-pVQZ. In all 26 different combinations of basis-function sets and levels of theory were utilised by \citet{hao}. 

In \S3.2 we compare our results with those of \citet{hao}. However, in this paper our emphasis differs greatly from theirs in that we have concerned ourselves with treating the large anharmonic corrections to the vibration frequencies of \hsix, and our study takes in isotopic variations for the reasons given earlier. We undertook detailed modelling of only the lowest energy configuration of the molecule, namely the $D_{2d}$ structure. We have, however, modelled the $D_{2h}$ transition structure at the harmonic level in order to gauge the effects of rotational delocalisation of the end-groups of the molecule (see \S3.2.1).

\subsection{Electronic excitations}
Excited electronic states of \hsix\ were investigated using the Equations Of Motion formalism \citep{krylov}.
In addition to the CCSD/cc-pVTZ level of theory we utilised CCSD/aug-cc-pVTZ (i.e. augmented by additional
diffuse basis functions) for this study. At the cc-pVTZ level we found the lowest two excited states to be
$2^2A_1$	 at 7.44~eV, and $1^2E$ at 8.79~eV. With the augmented basis these levels shifted to 7.43~eV
and 8.35~eV, respectively. The $2^2A_1$ state is reached by a valence transition and is unaffected by the additional
diffuse functions in the aug-cc-pVTZ basis.   The degenerate $1^2E$ state is stabilized by 0.44 eV in the
larger aug-cc-pVTZ basis, indicating that it is more diffuse. Not surprisingly, given that the binding energy
of the ground state of \hsix\ is only 2.1~eV \citep{kurosaki2}, the potential energy surfaces of both these
states are repulsive, so the molecule would dissociate if excited into either.

We have also investigated the influence of solvation on electronic excitations by computing the energies of
the two lowest-lying states for ${\rm H_6^+(H_2)_4}$. We found them to be substantially lower 
at  6.52 and 7.01~eV (EOM-CCSD/cc-pVTZ), but still repulsive. In practice we expect much higher levels of
solvation. For example, \citet{jaksch} found evidence for closed icosahedral solvation shells -- the first
closed shell corresponding to ${\rm H_6^+(H_2)_{12}}$, and the second to ${\rm H_6^+(H_2)_{12}(H_2)_{42}}$
-- when studying very cold (0.4~K) solvated clusters. We therefore expect that the lowest-lying electronic
excitations of the actual solvated form will be much lower than those of ${\rm H_6^+(H_2)_4}$ and
{\it we assume that some of these states will be bound states.\/} This assumption, and the change in the
electronic spectrum, on solvation, is important in the context of excitation of the vibrational modes of \hsix\ 
(see \S3.2.2).

Because we are dealing with a cation, we expect the energy required for any further ionisation to be large.
We have confirmed this by direct calculation: the energy required for photo-ionisation (``vertical ionisation'')
of bare \hsix\ is 19.2~eV, and that of ${\rm H_6^+(H_2)_4}$ is 18.3~eV.

\subsection{Vibrational model}
When computing the vibrational properties of a molecule the harmonic approximation is usually employed.
Such calculations are routine with the Q-Chem software package and yield both the frequencies of the normal
modes and the electric-dipole intensities of $|\Delta v|=1$ vibrational transitions. In the harmonic approximation
the Potential Energy Surface (PES) is taken to be quadratic about the equilibrium geometry. But \hsix\ is a weakly
bound system \citep{kurosaki2}, for which the higher-order terms in the PES are expected to be relatively large.
And furthermore the molecule is composed entirely of hydrogens, which are the lightest nuclei and thus have
the largest vibration amplitude in any given potential, so the higher-order terms in the PES are of particular
importance.

Our model of the effective PES, which we denote $U_{\rm eff}$, is a Taylor expansion  in the normal
mode coordinates, $q_i$,  of the system, around the potential minimum. For $N=6$ atoms there are $3N-6=12$ such
coordinates (the other six degrees of freedom are taken up by rigid body rotations and translations).
Our model includes the usual quadratic
terms, plus contributions that are cubic and quartic in the normal mode coordinates. In terms of the potential
derivatives, $\eta$ (the number of subscripts indicating the order of the derivative), we thus have
\begin{eqnarray}
\label{Ueff}
U_{\R{eff}}(\B{q}) & = &   \frac{1}{2!}\sum_{i}^{3N-6}\eta_{ii}q_iq_i
                     + \frac{1}{3!}\sum_{ijk}^{3N-6}\eta_{ijk}q_iq_jq_k \nonumber \\
                      &&+ \frac{1}{4!}\sum_{ijkl}^{3N-6}\eta_{ijkl}q_iq_jq_kq_l.
\end{eqnarray}
At the CCSD level only analytic gradients are available, so all the $\eta$ terms in equation (\ref{Ueff}) had to be computed
using finite difference methods to evaluate curvatures and higher-order derivatives of $U_{\rm eff}$. We used a step size,
$h$, of $0.1$~bohr and accurate finite difference expressions that give truncation errors of $O(h^4)$.

Anharmonic corrections to the vibration frequencies were computed using the Vibrational Configuration Interaction (VCI)
method \citep{whitehead,lin} in which up to five quanta were included.  For computational reasons only two-mode coupling was
considered. That is $\eta_{iij}$ terms were considered, but  $\eta_{ijk}$ terms, where $i\ne j\ne k$, were not. We also
excluded all couplings to vibration mode \#1 because, as discussed later (\S3.2.1), this mode is better characterised as
an internal rotation than a vibration.

\begin{deluxetable}{ccccrrr}
\tabletypesize{\scriptsize}
\tablecaption{Normal modes of the $D_{2d}$ structure of \hsix\ and \hd\label{table1}}
\tablewidth{0pt}
\tablehead{
\colhead{Species} & \colhead{} & \colhead{Mode} & \colhead{Symmetry} & \colhead{$I_o$} & \colhead{$\omega_o$} & \colhead{$\nu$} \\
\colhead{} & \colhead{} & \colhead{\#} & \colhead{ } & ${\rm(km\; mol^{-1})}$ & ${\rm(cm^{-1})}$ & ${\rm(cm^{-1})}$\\
}
\startdata
\hsix\  &   &   12  &   ${\rm   A_1}$   &   0.0\ \ \ \ \  &   3858.1  &   3596.0  \\
    &   &   11  &   ${\rm   B_2}$   &   953.3\ \ \ \ \    &   3779.1  &   3347.6  \\
    &   &   10  &   ${\rm   A_1}$   &   0.0\ \ \ \ \  &   2087.4  &   1816.0  \\
    &   &   ${}\;\;\;\,9^*$   &   ${\rm   E_{\ }}$ &   6.9\ \ \ \ \  &   1189.5  &   1104.8  \\
    &   &   ${}\;\;\;\,8^*$   &   ${\rm   E_{\ }}$ &   6.9\ \ \ \ \  &   1189.4  &   1104.7  \\
    &   &   ${}\;\,7$   &   ${\rm   B_2}$   &   2418.6\ \ \ \ \   &   1018.1  &   1042.1  \\
    &   &   ${}\;\,6$   &   ${\rm   A_1}$   &   0.0\ \ \ \ \  &   906.9   &   884.9   \\
    &   &   ${}\;\;\;\,5^*$   &   ${\rm   E_{\ }}$ &   13.6\ \ \ \ \     &   726.0   &   795.0   \\
    &   &   ${}\;\;\;\,4^*$   &   ${\rm   E_{\ }}$ &   13.6\ \ \ \ \     &   726.0   &   794.6   \\
    &   &   ${}\;\;\;\,3^*$   &   ${\rm   E_{\ }}$ &   0.3\ \ \ \ \  &   356.2   &   794.6   \\
    &   &   ${}\;\;\;\,2^*$   &   ${\rm   E_{\ }}$ &   0.3\ \ \ \ \  &   356.2   &   795.0   \\
    &   &   ${}\;\;\;\,1^*$   &   ${\rm   B_1}$   &   0.0\ \ \ \ \  &   99.2    &   201.3   \\
 & \\
\hd\ & & 12 &  &  6.4\ \ \ \ \ & 3343.3 &  3070.2\\
    & & 11 &  & 714.3\ \ \ \ \  & 3275.3 &  2947.7 \\
    & & 10 &  & 274.7\ \ \ \ \  & 1827.3 &  1574.2\\
    & &  ${}\;\;\;\,9^*$ &  &  5.4\ \ \ \ \  & 1073.1 &   962.9\\
    & &  ${}\;\;\;\,8^*$ &  &  47.6\ \ \ \ \  &  999.1  &  908.6\\
    & &  ${}\;\,7$ & & 1142.1\ \ \ \ \  &  865.1  &  832.5 \\
    & &  ${}\;\,6$ &  & 299.2\ \ \ \ \  &  800.7  &  738.1 \\
    & &  ${}\;\;\;\,5^*$ &  & 72.9\ \ \ \ \  &  616.8  &  626.6 \\
    & &  ${}\;\;\;\,4^*$ &  & 25.0\ \ \ \ \  &  557.4  &  613.8 \\
    & &  ${}\;\;\;\,3^*$ &  & 2.9\ \ \ \ \  &  297.8  &  225.3 \\
    & &  ${}\;\;\;\,2^*$ &  & 1.2\ \ \ \ \  &  291.0  &  184.6 \\
    & &  ${}\;\;\;\,1^*$ &  &  4.4\ \ \ \ \ &   84.7  &  120.5 \\
\enddata
\tablecomments{Wavenumbers are given both with ($\nu$) and without
($\omega_o$) anharmonic corrections. Changes in rotational state of the molecule shift
the lines away from these wavenumbers. Modes marked with an asterisk are strongly
affected by rotation of the end-groups, so these modes
are not good approximations to the eigenmodes of the real molecule (\S3.2.1).}
\end{deluxetable}

Table 1 shows our results for the normal vibrational modes of \hsix. There we list the harmonic frequency,
the electric-dipole intensity (derived in the harmonic approximation), and the anharmonic frequency for each
mode. The four intensities listed as zero are, in the harmonic approximation, precisely zero in consequence
of the symmetry of the mode (${\rm A_1}$ or ${\rm B_1}$). Of the remaining eight modes,  all six of E symmetry are
fairly quiet while the two ${\rm B_2}$ modes are both strong. These two modes acquire their intensity as a result of
displacing the central H$_2^+$ moiety along its axis.
Indeed mode 7 can be thought of as the central H$_2^+$ moiety bouncing back and forth along the main
axis of the molecule and this mode is by far the most intense. Modes 12 and 11 both involve stretching of the two \htwo\
end-moieties and axial displacement of the central atoms. The two modes differ in that mode 12
is a symmetric manifestation of these motions whereas mode 11 is antisymmetric. Consequently the latter involves a
displacement of the central H$_2^+$ moiety and thus acquires a large intensity, whereas mode 12 simply stretches
this moiety. Figure 2 shows the motions involved in each of the normal modes of \hsix.

\begin{figure}
\figscaleone
\plotone{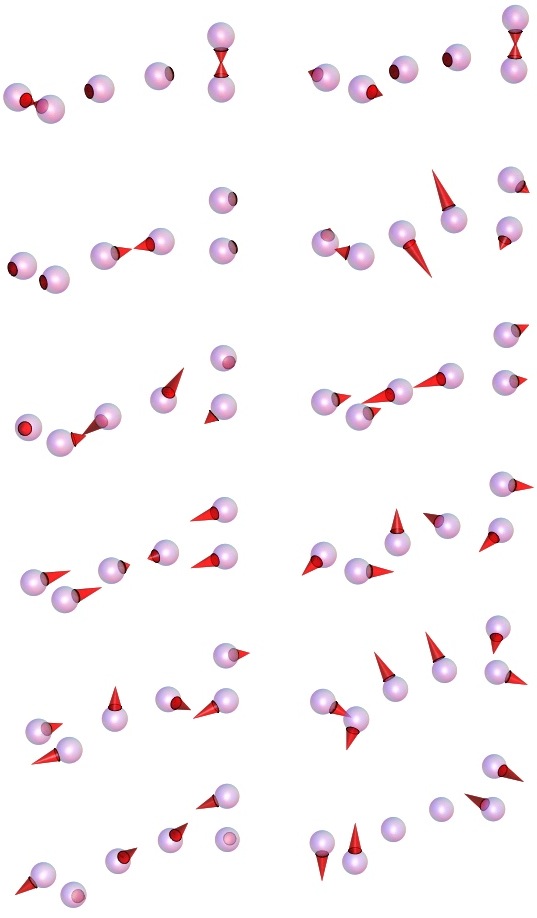}
\caption{The normal modes of vibration of \hsix. The left column shows the even-numbered modes, in descending order starting from mode \#12 at the top. The right column shows the odd-numbered modes, in descending order starting from mode \#11 at the top. Spheres show the locations of the atoms at rest and red cones show the amplitude and direction of motion for each atom.}
\end{figure}

Comparing the frequencies in the two rightmost columns of table 1 shows clearly the high degree of anharmonicity
in this molecule, exceeding 400\percm\ for some modes. 

Having obtained the PES model as per equation (\ref{Ueff}) most of the computational work associated with the vibrational
calculations is done and it is straightforward to investigate the isotopic variants of \hsix. In \S2.5 we noted
that condensation of HD onto \hsix\ is expected to proceed; so in table 1 we also
present the normal modes of \hd. For this isotopomer the vibrational symmetries of \hsix\ are broken
and thus there are no silent modes. Indeed modes 10 and 6 acquire large
intensities. Because the effective oscillating mass is greater for \hd\ than for \hsix, each of the normal modes of \hd\ 
has a lower frequency than the corresponding mode for \hsix. And whereas \hsix\ shows only two strong lines (in the harmonic approximation),
there are at least four strong lines for \hd. The lines for both isotopomers are shown in figure 3.

The normal modes of vibration are not the whole story. In the harmonic approximation the normal modes are the
only modes and they are non-interacting. The intensities are the same for all $|\Delta v|=1$ transitions of a given
mode, and zero for $|\Delta v|\ge2$. But we have seen that the anharmonicity is large for \hsix\  and we therefore
expect that there may be significant intensities associated with some of the overtones ($|\Delta v|\ge2$) and combination bands
(which result from coupling between the normal modes). Although we have calculated the frequency shifts due to
anharmonicity, we are unable to compute the intensities of overtones or combinations.

\subsubsection{Errors and uncertainties}
There are various sources of error in our vibrational model of \hsix; although we cannot remove these errors
it is useful to know roughly how large they are. The easiest check to make is whether or not the predicted
harmonic properties of \hsix\ vary significantly with the level of electronic structure theory which is utilised.
That test can be made by examining the results of \citet{hao}, either in isolation or by comparison with our own
results. The highest levels of theory utilised by \citet{hao}
are cc-pVTZ UCCSDT and aug-cc-pVQZ UCCSD(T) and at these levels the harmonic frequencies typically differ
by amounts (mean absolute difference) of order 11\percm\  and 17\percm, respectively, compared to our own results.
These differences constitute a small fractional error for most of the vibrational modes.

\begin{figure}
\figscaleone
\plotone{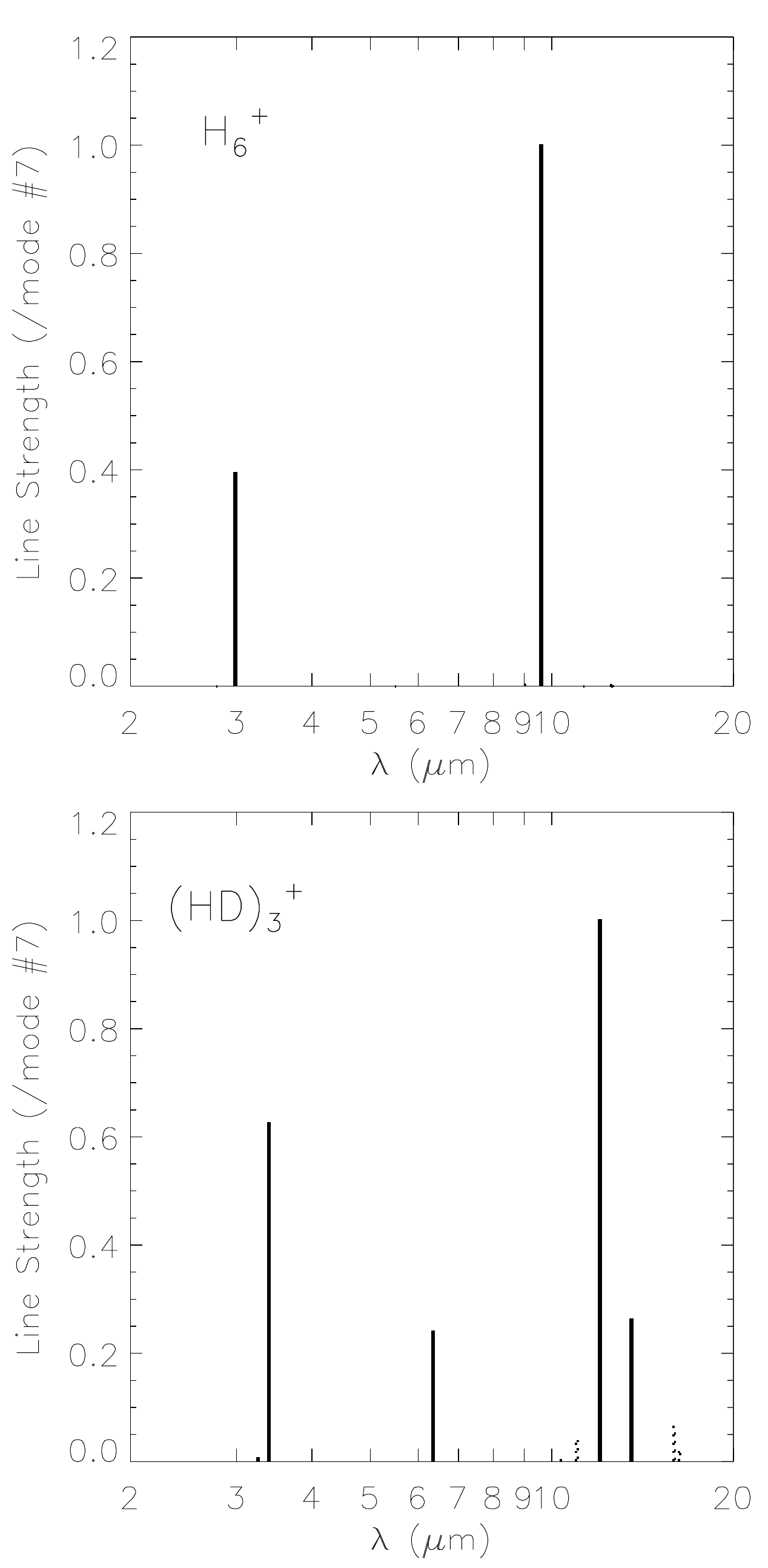}
\caption{Stick diagram of the mid-IR lines of \hsix\ (upper panel), and \hd\ (lower panel). The line strengths are normalised
to that of mode \#7 for each molecule. The E-type modes of the $D_{2d}$ structure are not good approximations to the
actual eigenmodes of the molecules (see \S3.2.1) and they are therefore plotted with dotted lines.}
\end{figure}

Our level of theory corresponds precisely to the cc-pVTZ UCCSD case of \citet{hao}. In this case the mean absolute
frequency difference between the two sets of results is only 6\percm\  (or 4\percm\ if mode \#1 is excluded from the
comparison).  We attribute these differences to the fact that \citet{hao} use analytic derivatives whereas we employ
numerical derivatives.

In respect of the predicted (non-zero) intensities of the various modes there is, however,
a substantial difference between our results and those of \citet{hao}. Comparing our results to the highest levels
of theory employed by \citet{hao} we see good agreement only for mode \#11 (difference less than 10\%), whereas
for mode \#7 their predicted intensity is less than half of ours and for modes \#8 and 9 it is larger by a factor of 10.
The disagreement is larger still for the four remaining modes, their intensities being more than 20 times larger
for \#4 and 5, and more than one thousand times larger for \#2 and 3. However, these discrepancies parallel
those which are evident amongst the various levels of theory of \cite{hao} for each of the modes, to wit:
discrepancies of order 10\% for mode \#11; a range spanning a factor of 3 for modes \#7, 8 and 9; a factor
of 30 for modes \#4 and 5; and a factor of 100 for modes \#2 and 3. 

Given the huge variation in intensities reported by \citet{hao}, for the same mode with different levels of theory,
we have no cause to doubt our own estimates of the mode intensities.

By comparing our calculations to observed vibrational characteristics we are also able to gauge
errors beyond just the harmonic level. These checks cannot be made against \hsix\ itself, because
the vibrational frequencies of that molecule have not yet been measured. But we have checked the accuracy
of the quartic approximation in equation (\ref{Ueff}) for the case of the \htwo\ molecule: we find that the computed vibration
frequency is 61\percm\ above the measured value for that molecule. Employing a sextic approximation to the
PES reduces this error to only 2\percm, so the error associated with the quartic approximation of equation (\ref{Ueff})
appears to be roughly 60\percm\  (at 4,159\percm). Another major source of error in our calculations of
vibrational frequencies stems from our use of only two-mode coupling in the VCI method. By modelling
${\rm H_3^+}$ we find that the errors introduced by restricting our calculations to two-mode coupling could
be as large as 50\percm (at 2,521\percm).

All of our results here pertain to the isolated \hsix\ molecule, and its isotopomers, whereas we expect that \hsix\
inside a matrix of solid \htwo\ would be in the solvated form, which will have slightly different vibrational properties.
In \S3.4 we give quantitative estimates of the effects of solvation on the vibration frequencies of \hsix\ by noting the
shifts which occur at the harmonic level between \hsix\ and ${\rm H_6^+(H_2)_4}$.

\begin{figure}
\figscaleone
\plotone{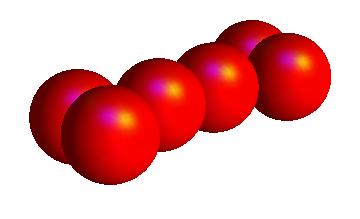}
\caption{The $D_{2h}$ transition structure of \hsix\ differs from the $D_{2d}$ structure of figure 1 only in
respect of the dihedral angle between the end-groups. This structure is a saddle point of the PES and
it lies only 12 \percm\  above the $D_{2d}$ minimum \citep{kakizaki}.}
\end{figure}

Finally we draw attention to a fundamental source of uncertainty in our modelling of the vibrations of \hsix.  Although the
$D_{2d}$ structure is the minimum energy configuration of \hsix, the $D_{2h}$ structure, in which the two end-groups
lie in the same plane (see figure 4), is a transition structure which lies only 1.5~meV (12\percm) above the minimum
\citep{kakizaki}. This transition structure can be reached simply by counter-rotation of the end-groups of \hsix, i.e.
the degree of freedom manifest in vibration mode \#1 (see figure 2). But the vibration frequency for mode \#1 is 
larger than the barrier height and even the zero-point vibration energy is sufficient to completely delocalise the
structure in this coordinate. The resulting probability distribution functions have been computed by \citet{kakizaki}
for the isotopomers \hsix\ and ${\rm D_6^+}$ and in both cases they are almost uniform in the dihedral angle.
The situation is more complicated in the case of \hd, because the centre-of-mass of the HD end-groups does
not coincide with the mid-point of the bond, around which the groups rotate. For small amplitude vibrations
(in mode \#1) the result is simply that the H atom oscillates with twice the amplitude of the D atom. But that cannot
be the case for large amplitude motions: if one atom performs a complete rotation, so must the other. Rather we
anticipate that the atoms of the central HD$^+$ group will move in response to the rotation of the end-groups.
The \hd\ isotopomer was not modelled by \citet{kakizaki} but we assume that, like \hsix\ and ${\rm D_6^+}$, the
end groups of the molecule are delocalised in the dihedral angle.

\begin{deluxetable}{ccrrrr}
\tabletypesize{\scriptsize}
\tablecaption{Vibrational  frequency shifts for $D_{2h}$ structure and solvated form of $D_{2d}$.\label{table2}}
\tablewidth{0pt}
\tablehead{
\colhead{\ \ Mode\ \ } & \colhead{Symmetry} &\colhead{$\Delta_{dh}$} & \colhead{}&\colhead{$\Delta_{s}$}  \\
\colhead{\#} &  \colhead{ } & ${\rm(cm^{-1})}$ & \colhead{}& ${\rm(cm^{-1})}$ \\
}
\startdata
12  &     ${\rm   A_{1}}$      &  6    &    &$-$50    \\
11  &    ${\rm   B_{2}}$    &  3    &    &$-$68     \\
10  &    ${\rm   A_{1}}$       &  10    &    &$-$23     \\
9    &    ${\rm   E_{\ }}$       &  55   &   &5    \\
8    &   ${\rm   E_{\ }}$    &$-$115 &   &5     \\
7   &      ${\rm   B_{2}}$     &$-$10   &   &25      \\
6   &     ${\rm   A_{1}}$      &$-$3  &    &27       \\
5   &     ${\rm   E_{\ }}$       &120   &   &34         \\
4   &    ${\rm   E_{\ }}$    &$-$112  &    &34        \\
3   &    ${\rm   E_{\ }}$       &39  &    &  56              \\
2   &    ${\rm   E_{\ }}$       &$-$17  &    &  56            \\
1   &     ${\rm   B_{1}}$     &  -  &  &  -                     \\
\enddata
\tablecomments{The symmetry designation refers to the isotopomer  \hsix\ in the $D_{2d}$ structure.
Frequencies appropriate to the $D_{2h}$ structure are given by adding $\Delta_{dh}$ to the
values in Table 1.  Frequencies appropriate to the solvated species are
given by adding $\Delta_{s}$ to the values in Table 1.}
\end{deluxetable}

Delocalisation affects the vibrational characteristics of the molecule. Strictly speaking it means that none
of our analysis is valid, because $U_{\R{eff}}$ is a sinusoidal function of $q_1$ which, over
its full range, is very poorly represented by equation \ref{Ueff}. And to the extent that the properties of a vibration
mode depend on the dihedral angle that mode is necessarily coupled to mode \#1. All couplings involving mode
\#1 are explicitly excluded from our model because mode \#1 has more the character of an internal rotation than
a vibration.

To gauge the size of these effects we can compute the harmonic vibration frequencies for the $D_{2h}$ structure
and compare them to those of the $D_{2d}$ structure. We have done this and in table 2 we show our results as the
quantity $\Delta_{dh}=\omega_o(D_{2h})-\omega_o(D_{2d})$. 

For mode \#1 the $D_{2h}$ structure has an imaginary
vibration frequency  because it is a maximum of the PES traversed along the coordinate $q_1$. For modes \#6,7,10,11 and 12
we find that the frequencies are the same as those of the $D_{2d}$ structure to within 10\percm\ and the mean absolute
difference is only 6.4\percm. These shifts are very small and that tells us that those modes are good approximations
to the eigenmodes of the actual system, with its rotationally delocalised end-groups. We can therefore be confident
that we have a reliable estimate of the frequencies of the corresponding lines.

But for the remaining six modes, all of which have E-type symmetry, the shifts are an order of magnitude greater.
Degeneracy of the three pairs of E-modes of \hsix\  in the $D_{2d}$ structure is lifted in $D_{2h}$ with splittings of
56\percm\ (modes \#2,3), 232\percm\ (modes \#4,5) and $170$\percm (modes \#8,9). Thus the E-modes are
strongly influenced by the nearly-free rotation of the end-groups of the \hsix\ molecule.

The reason for this striking difference
between the E-modes and the modes of A and B symmetry is that the E-modes all involve displacements of the
central two atoms perpendicular to the $z$-axis (i.e. the long axis of the molecule), 
whereas the other modes only displace these atoms along the $z$-axis (see figure 2). Thus E-mode vibrations are
affected by the orientation of the end-groups, but the other modes are not.

The large frequency shifts of the E-type modes between $D_{2d}$ and $D_{2h}$ structures indicate that the
corresponding vibrational eigenstates which we have characterised for the $D_{2d}$ structure are poor approximations
of the actual eigenstates of the rotationally delocalised molecule. We therefore place correspondingly little
weight on these modes when comparing our results to observational data. 

\subsubsection{Excitation of vibrations}
In solid \htwo\ the temperature is necessarily below the triple-point of hydrogen so we have $kT<10$\percm,
whereas the smallest vibrational quantum of \hsix\ exceeds 100\percm, thus all of its excited vibrational states
lie too far above the ground state to have any significant thermal population. There will thus be no thermal
emission of vibrational quanta from \hsix\ in solid \htwo.  But the vibrational modes of \hsix\ can be excited by
non-thermal interactions. The most important non-thermal process is likely to be absorption of ambient
starlight, for which the energy density peaks at a wavelength of approximately 1~$\mu$m \citep{mathis},
corresponding to photon energies $\sim1$~eV.

There are two mechanisms by which \hsix\ may absorb ambient starlight: via electronic excitation and
through vibrational combination bands. Ordinarily one would expect the former to be by far the more
important of these two processes, as combination bands are usually very weak. However, there are
three aspects of \hsix\ which suggest that combination bands may not be completely negligible. First
is that the energy of the lowest excited electronic state is very high at 7.44~eV, where the energy
density of the ambient radiation field is only one fifth to one tenth of its value at the peak \citep{mathis}.
By contrast there are many vibrational combination bands of \hsix\ distributed right across the broad spectral
peak of the ambient starlight.  Secondly, even the lowest excited electronic state lies well above the vibrational
continuum of the electronic ground state of \hsix, which has a binding energy of only 49~${\rm kcal\;mol^{-1}}$,
or 2.1~eV per molecule \citep{kurosaki2}. This means that the bulk of the photon energy cannot go into exciting
the vibrational modes of \hsix. Indeed we expect that there will be essentially no excitation of vibrations as the two lowest
excited electronic states are repulsive and dissociation should occur promptly. Thus electronic excitation is an
inefficient way of powering vibrational emission from \hsix. This is in stark contrast with excitation through
vibrational combination bands --- a process which is 100\% efficient. Finally we note that \hsix\ is an unusually
floppy, anharmonic molecule and hence we expect it to have unusually strong combination bands. 

Solvated forms of \hsix\ are also floppy and thus are also expected to have unusually strong combination bands.
However, we have found (\S3.1)  that solvation to ${\rm H_6^+(H_2)_4}$ lowers the minimum electronic excitation
energy significantly. And for greater degrees of solvation we expect the energy to decrease further; we {\it assume\/}
that this gives rise to excited electronic states which are bound. Lowering the energy of the excited state
increases the efficiency of excitation of the vibrational
degrees of freedom and it brings the electronic excitations into the spectral peak of the ambient starlight. We therefore
anticipate that for highly solvated forms of \hsix, which are prevalent at low temperatures, electronic excitation is the dominant
process.

Unfortunately we cannot sensibly model the electronic excitation of vibrations, because the excited electronic states
are sensitive to the form of solvation of the molecule. Consequently we are currently unable to model the optical
absorption spectrum of \hsix\ or the vibrational level population subsequent to such absorption. It is the latter which
determines the strength of any vibrational emission because the cascade of emitted quanta will be dominated by
$\Delta v=-1$ transitions in the various modes which have been excited.
{\it Thus we are unable to model the intensities of the mid-IR emission lines from \hsix.\/} 

We now consider the band structure which arises from rotational transitions of the molecule during emission
of a vibrational quantum.

\subsection{Rotational band structure}
In the rigid-body approximation the rotational properties of \hsix\ are characterised by its moments of inertia or,
equivalently, its rotational constants. For \hsix\ these are $B_x,B_y=1.38$\percm\ and $B_z=26.8$\percm\ and
for \hd\ $B_x,B_y=0.92$\percm\ and $B_z=18.5$\percm. Thus both isotopomers resemble a prolate symmetric top.
Because $B_x,B_y\ll kT < B_z$ in both cases, rotation about the $x$ and $y$ axes may be thermally excited,
whereas there will be little thermal excitation of rotation around $z$.

As emphasised in \S3.2.1, \hsix\ is actually a floppy molecule and rigid-body rotation results are therefore not a good
guide to its rotational properties.

Solvated species of \hsix\ will also be floppy because the \hsix\ itself is only slightly perturbed by solvation and the
bonds between the solvent and solvated molecules are themselves weak. The impediment to calculation therefore
exists also in this case. A further uncertainty for the solvated species is whether or not such a large complex is able to rotate at all inside a
matrix of solid hydrogen. Alternatively, internal rotation of the \hsix\ might occur relative to the solvating shells.

{\it In view of the difficulties just described we are not able to undertake sensible calculations of the rotational band
structure for \hsix.\/}

\subsection{Model spectra}
As we are unable to model either the strength of the vibrational emission or the band structure
appropriate to the various lines, the only real information contained in our model spectra is the line frequencies.
We therefore present the results of our modelling simply by marking the wavelengths of the vibration modes in figure 5.
Results are shown for both \hsix\ and \hd. 

We remind readers that modes of E symmetry are
not good approximations to the eigenmodes of the real molecule; they are therefore shown with dotted
lines, to draw attention to the unreliable nature of the prediction. We further emphasise the effect of the
rotation of the end-groups by plotting each mode as a box whose spectral width spans the frequencies predicted
for the $D_{2d}$ and $D_{2h}$ structures (see table 2). For the five well characterised modes these boxes are
sufficiently narrow that they resemble vertical lines in figure 5.

\subsubsection{Effects of solvation}
Following \citet{kurosaki2} we have characterised the harmonic vibration frequencies of the solvated molecule
${\rm H_{14}^+=H_6^+(H_2)_4}$ in $D_{2d}$ symmetry. Because of the much larger number of atoms,
${\rm H_{14}^+}$ inevitably has many more modes of vibration --- 24 more than \hsix\ in fact. Of these the
highest four frequencies are associated with stretching of the four solvent molecules, and most of the lowest
frequencies are associated with vibrations which involve the weak bonds between these molecules and the
core \hsix. In between there are modes which can be readily identified as perturbed modes \#2--12 of the core
\hsix. Comparing the frequencies of these modes with the harmonic frequencies for bare \hsix\ (table 1) gives 
us an indication of the likely shifts associated with solvation. Our results are given as the quantity $\Delta_s$
in table 2 --- a frequency shift which should be added to the predictions for the bare molecule in order to obtain
the vibration frequencies of the solvated form.

We now apply these shifts to our {\it anharmonic\/} frequency predictions for \hsix\ and
\hd\ in order to estimate the vibration frequencies for the solvated forms of these molecules. The results of this
procedure are shown in figure 5.

\bigskip\goodbreak

\section{\hsix\ spectra vs astronomical data}
In addition to our theoretical predictions, we have shown in figure 5 the ISO mid-IR spectrum of
the Orion Bar \citep{peeters2}.  This spectrum was chosen because the data are of high quality and
the band strengths and profiles are representative of the sort of spectrum which is most commonly
seen, i.e. the ``Type A'' spectrum \citep{geballe,tokunaga,peeters1,diedenhoven}.

For clarity of presentation we have eliminated from this plot sixteen narrow lines arising from atomic
or molecular hydrogen, neon, sulphur or argon. The lines are \citep{vandishoeck}: H, Brackett-$\alpha$;
four lines from the Pfund series of H; five pure rotational lines from \htwo; [Ar II]~7.0\mum; [Ar III]~9.0\mum;
[S IV]~10.5\mum; [Ne II] 12.8\mum; [Ne III] 15.6\mum; and [S III] 18.7\mum.  We thus concentrate on
the broad bands of observed emission \citep{gillett}; these are often referred to as the Unidentified InfraRed
(UIR) bands  and we use that term henceforth.

It has been proposed that the UIR bands are vibrations of Polycyclic Aromatic Hydrocarbons (PAHs)
\citep{duley,leger,allamandola} and this suggestion has been examined in considerable detail --- see
the review by \citet{tielens}. However, the identification with PAHs is not completely secure.  In part that
is because there is no predictive theory for the shapes or relative strengths of the observed bands.
This means that if the emission from \hsix\ lies within these bands then it will be
difficult to place tight constraints on the abundance of the isotopomer in question.

\begin{figure}
\figscaleone
\plotone{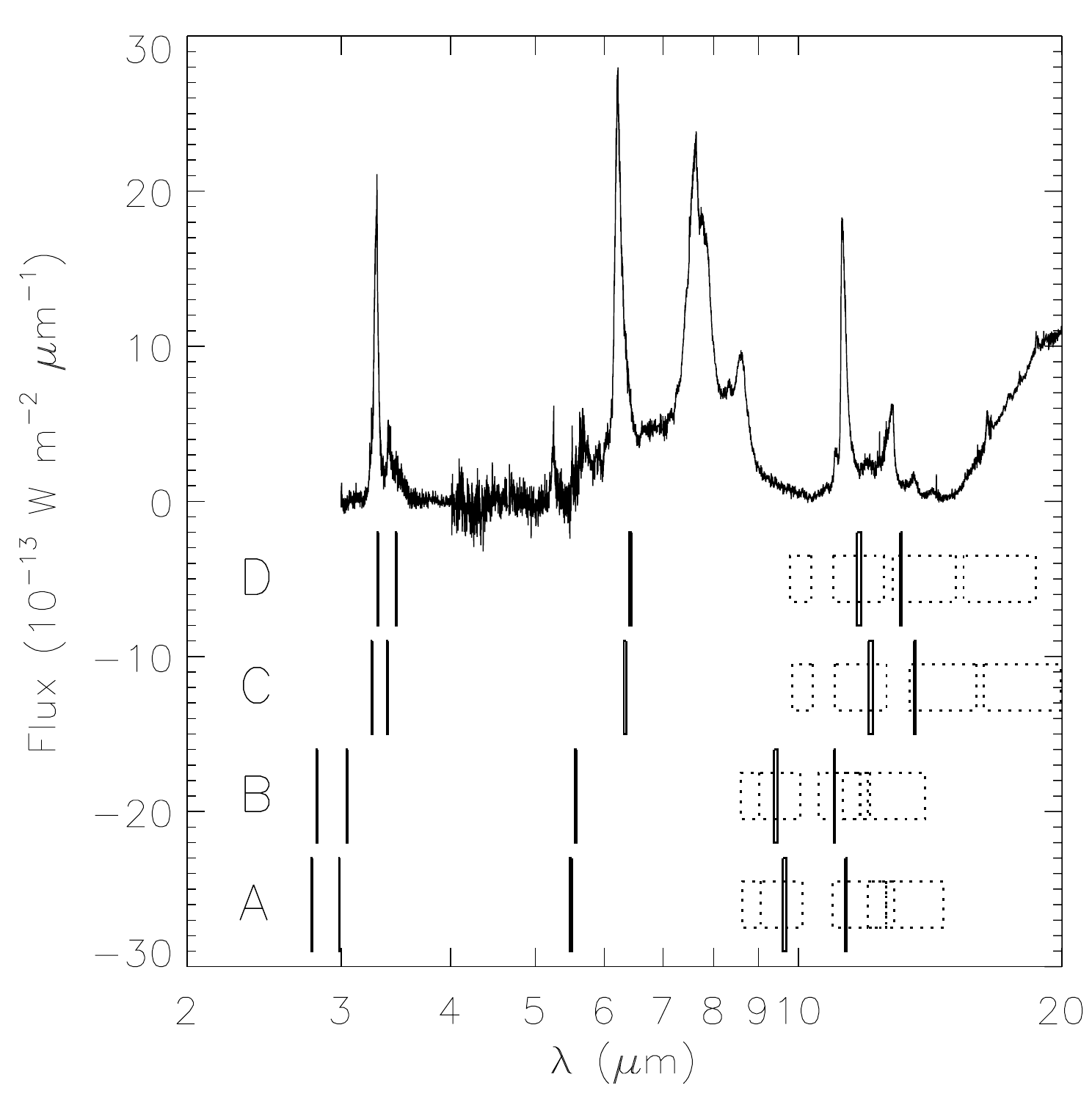}
\caption{Infrared Space Observatory spectrum of the Orion Bar \citep{peeters2}. A background flux $\propto\lambda^{3.2}$ 
has been subtracted from the data so that the broad emission bands can be seen more easily. Also plotted here,
in the lower half of the figure, are the locations of the predicted fundamental vibration lines for \hsix\ and \hd\ 
isotopomers in bare and solvated forms. Each line is stretched into a band whose spectral width spans the predictions
for $D_{2d}$ and $D_{2h}$ structures in the case of the bare isotopomers. Reliable predictions are shown with solid
lines whereas the E-type modes are indicated with dotted lines. Predictions are coded thus: $A\leftarrow{\rm H_6^+}$;
$B\leftarrow{\rm H_6^+(H_2)_4}$; $C\leftarrow{\rm (HD)_3^+}$; and $D\leftarrow{\rm (HD)_3^+(H_2)_4}$.}
\end{figure}

First we compare the Orion spectrum with our results for the \hsix\ isotopomer, restricting attention to
the five lines whose frequencies we can predict with confidence. Overall we see no clear correlation
between the data and our predictions for \hsix, in either bare or solvated forms. For bare \hsix\ mode
\#6 lies right on the strong 11.3\mum\ (885\percm) band, and for the solvated molecule this mode is
less than 30\percm\ from the band, so within the possible errors (see \S3.2.1) both species yield a coincidence.
It is more difficult to claim a coincidence with any of the other modes, either for solvated or bare \hsix.
Unfortunately the data shown here do not extend shortward of 3\mum; however, we are not aware of
any strong emission bands in this region. The closest feature to modes \#11 and 12 is thus the
3.3\mum\ emission band seen in the plot and this is so far removed from the predicted lines that it
is plainly not attributable to the \hsix\ isotopomer in either bare or solvated forms. Similarly, mode
\#10 is almost 200\percm\ away from the nearest strong emission band at 6.2\mum.

Translating the lack of spectral coincidences into limits on the \hsix\ abundance requires
information which we do not yet have. However, it is fair to say that the
data in figure 5 give no reason to think that the \hsix\ isotopomer is abundant in interstellar space.
Although the data shown in that figure are an example of a ``Type A'' spectrum, we would have
reached the same conclusion if we had compared our results to a Type B or Type C spectrum
as these differ from Type A mainly in respect of the relative strengths and shapes of the various bands,
rather than in their placement \citep{peeters1,diedenhoven}. It remains possible that spectral coincidences
could be found in other sorts of data --- see the discussion of absorption spectra in \S5.3.

The situation is quite different for \hd: figure 5 shows a striking coincidence between the data and our
predictions, especially for the solvated form. Coincidences are seen in the case of the emission bands
at 3.3\mum\ (mode \#12), 3.4\mum\ (mode \#11), 6.2\mum\ (mode \#10), 11.3\mum\ (mode \#7), and
12.7\mum\  (mode \#6). And we emphasise that this accounts for all of the reliable mode predictions for
\hd.  There is no predicted mode coincident with the strong emission bands at 7.7 and 8.6\mum, but this
is of no immediate significance since the remaining vibration modes, the E-type modes, are not good
approximations to the eigenmodes of the actual \hd\ molecule (\S3.2.1). Notwithstanding the uncertain
identification of the 7.7 and 8.6\mum\ emission bands, the excellent agreement between the data and
the well characterised modes of \hd\ prompts us to suggest that this molecule is responsible for the
ubiquitous UIR bands.

\bigskip\bigskip

\goodbreak
\section{Discussion}
Having suggested that \hd\ is the carrier of the UIR bands we now need to consider the broader
implications of that suggestion. For example, the UIR bands typically carry a significant fraction
($\sim10$\%) of the total intensity seen on various lines of sight through our Galaxy \citep{onaka}, so
their carrier must be contributing substantially to the extinction on these lines of sight. In this section
we consider \hd\ in the broader picture of ISM photo-processes. Readers who are unfamiliar with
current models of interstellar dust may wish to consult \citet{draine2} and the many papers in
\cite{wittclaytondraine}. The following discussion is couched entirely in terms of \hd, even though
the same physics would in most cases apply to the \hsix\ isotopomer.

\subsection{Excitation mechanism}
In \S3.2.2 we argued that excitation of the vibrational modes of \hd\ is expected to occur as a result
of absorption of ambient starlight. The two contributing processes were identified as electronic
excitation and absorption in combination bands, with electronic excitation expected to dominate for
highly solvated forms. Some insight into the excitation process is available from the data
in figure 5 where we see that there are peaks at both 3.3 and 3.4\mum\ with the former
being by far the stronger of the two. Our model of \hd\ does indeed predict lines at approximately
these locations -- modes \#12 and 11, respectively -- but the ratio of calculated transition strengths
is the opposite of that observed:  mode \#11 is more than one hundred times stronger than mode \#12.
This suggests that if the observed bands arise from \hd\  then the excitation process is probably electronic.
The reason is that the strength of a combination band in part reflects the strength of the individual transitions
involved in that combination, so if these modes were excited by absorption in combination bands then we
would expect that mode \#11 would be the stronger of the two lines. By contrast, if the excitation is electronic
then the population of the excited vibrational levels is determined by the Franck-Condon factors and is unrelated
to the transition strengths of the modes. It seems probable that the Franck-Condon factors would favour excitation
of mode \#12. The reason is that mode \#12 is a symmetric vibration mode, whereas mode \#11 is antisymmetric.
The Franck-Condon factors reflect the distortion in the molecular PES which occurs as a result of the electronic
excitation, and for the low-lying electronic excitations those distortions will be predominantly symmetric. 

\subsection{Relationship to FIR}
In this paper we have concentrated on the vibration modes of \hd\ itself. But the solvated form has in addition
a large number of vibration modes which we have thus far ignored. For example, in the case of icosahedral
solvation with a single shell, i.e. ${\rm (HD)_3^+(H_2)_{12}}$, we would expect a total of 84
modes, whereas \hd\  has only 12. Extrapolating from the behaviour seen for low degrees of solvation
\citep{kurosaki2} we expect that a complex with a single icosahedral shell would exhibit 12 high-frequency
modes, close to the vibration frequency of unperturbed \htwo, and 60 low-frequency modes, in addition to
the modes common to bare \hd. If solvation extends to a second complete icosahedral shell, i.e.
${\rm (HD)_3^+(H_2)_{12}(H_2)_{42}}$, then we expect there to be 270 low-frequency vibration modes
attributable to the solvation.

Following electronic excitation of the solvated complex, courtesy of an absorbed photon, it is probable
that many of the low-frequency modes will be excited because the lowest-lying electronic excitations
are likely to be spread over the whole complex. Thus if \hd\ is responsible for the UIR bands then we expect
that Far InfraRed (FIR) emission will accompany them. Because of the large number of modes crammed
into a relatively small emission bandwidth, the FIR emission may appear as a quasi-continuum if the degree
of solvation is high.

FIR emission is indeed observed to accompany UIR band emission and in fact there is generally
observed to be far more power in the FIR than in the UIR bands (e.g. Onaka 2004). In a model
which seeks to attribute the observed FIR power to solvated \hd\ that would imply that the majority
of the observed interstellar extinction must be attributed to absorption by \hd.

\subsection{Absorption spectra}
If \hd\ is present in interstellar space then it should manifest itself in absorption as well as emission.
The relative strengths of the infrared vibrational lines may differ greatly between these two processes.
Absorption-line strengths just reflect the oscillator strengths of the transitions (assuming the molecule
is in the vibrational ground state). These mode intensities are given in table 1
and plotted in figure 3. Thus we see, for example, that in absorption the 3.4\mum\ 
line of \hd\ is expected to be much stronger than the 3.3\mum\ line, whereas in emission the 3.3\mum\ band is
observed to be more intense. In practice it is unusual to see any of the UIR bands in absorption, so there isn't
much data to compare our results with. However, there is sometimes an aborption band seen at 3.4\mum\ 
with no detectable 3.3\mum\ absorption accompanying it \citep{pendleton}. Conventionally this
band is interpreted in terms of aliphatics --- long-chain hydrocarbon molecules, very different to the
polycyclic species which are usually invoked to explain the other UIR bands.

We are not aware of any other interstellar absorption bands coincident with our mode predictions for \hd.
However, as our predictions match the locations of the observed UIR emission bands it seems unlikely
that this situation presents any particular difficulty for interpreting the UIR emission in terms of \hd.

Looking again at figure 3, but this time at the predictions for the \hsix\ isotopomer, we see that
the only strong absorption lines expected for that molecule are at 9.6\mum\ and 3.0\mum\ --- coincident
with known bands of strong interstellar absorption. These bands are usually attributed to
amorphous silicates and to water ice, respectively \citep{draine2,vandishoeck}.  It is beyond the scope
of this paper to construct an astrophysical model in which these absorption bands are interpreted in
terms of \hsix; here we simply note the spectral coincidences.

We have already noted that the optical absorption spectrum of solvated \hsix\ is probably dominated
by electronic excitations (\S3.2.2) and that this expectation is reinforced by the data in figure 5 (\S5.1).
We are unable to calculate the electronic absorption spectrum, but we can give a qualitative
description as follows.

The large number of vibrational modes possessed by a single solvated molecule means that we
can expect to see a very large number of absorption lines. Taking the bare molecule, for example,
there are 12 modes of vibration. Assuming the Franck-Condon factors favour only two different
quantum states for each mode gives us $2^{12}\sim4,\!000$ different vibrational states and the same
number of absorption lines. For the solvated species this number is, of course, vastly greater: for a single
icosahedral shell, i.e. ${\rm (HD)_3^+(H_2)_{12}}$, it is $\sim10^{25}$.  Given that the binding energy
of the clusters is only about 2.2~eV in the ground electronic state \citep{kurosaki2}, the spacing between
individual lines is expected to be typically very small so that they create, in effect, a continuum absorption.

Individual absorption lines should nevertheless be detectable because the large number of
vibration modes also leads to a large dynamic range in line strength. For example, if we assume,
as before, that the Franck-Condon factors favour only two different quantum states in each mode,
but that one of these states has typically twice the Franck-Condon factor of the other, then the
intensity ratio between the strongest and weakest lines is of order the total number of lines.

All of the foregoing arguments apply equally well to any bound, excited electronic states.
It is likely that there will be a number of such states and this correspondingly multiplies the number of
possible absorption lines. 

An unusual aspect of \hd\ is its high levels of anharmonicity. One result
of this is that in any vibrational sequence the level spacings are not uniform. Thus, even in cases
where the Franck-Condon factor is large over a broad range of one vibrational quantum number,
there will be no regular spacing in the corresponding sequence of absorption lines.

If we try to make a connection to the astronomical data then we are looking for correlated optical continuum
and line absorption phenomena in which the optical lines are unidentified. We therefore consider the
possibility of a connection to the Diffuse Interstellar Bands (DIBs). The properties of DIBs have
been reviewed by \citet{herbig3} and \cite{sarre}, for example. Their phenomenology is rich and
we do not attempt a detailed assessment vis-\'a-vis \hd\ absorption; instead we confine ourselves
to the following points. First, DIBs are known to correlate well with optical continuum absorption,
as gauged by reddening. Secondly, DIBs are concentrated in the optical band, consistent with
a carrier that has a small binding energy. Thirdly, although there have been suggestions of
regularities in DIB spacings (notably, Herbig 1988), with the best available data it appears that
there are no statistically significant regularities in line spacing for intervals smaller than 400\percm\ 
 \citep{hobbs1,hobbs2}. This is consistent with absorption by a floppy molecule.
 
Finally, the lack of very strong correlations amongst the various DIB strengths, measured on different
lines of sight \citep{mccall,friedman}, is a puzzling feature of the data. But it can be understood in the context of
an absorption spectrum in which any individual absorption line represents a miniscule fraction of all
the absorptions, as we expect for solvated \hd. This is because that fraction will depend on the smallest
details of the wavefunctions of the ground- and excited-states and the dependencies will be different for
each absorption line. Branching ratios may therefore exhibit a strong environmental dependence.

We also note that DIB strength is known to correlate with the column-density of atomic hydrogen,
not molecular hydrogen \citep{herbig2,friedman}. At first sight this might seem to argue against \hd\ being the carrier,
given its intimate association with solid molecular hydrogen. However, the correlations are
determined using measurements that are sensitive to diffuse gas and insensitive to precipitates.
Furthermore it is possible that the diffuse atomic hydrogen could actually arise from photodissociation
of \htwo\ which is itself undetected \citep{allen}.

Finally we recall that \hd\ resides in a matrix of solid \htwo, which contributes absorption, as do the
\hminus\ ions which we expect are also present (\S2.5). The electronic transitions which make up the Lyman and
Werner bands of \htwo\ are very strong and they are expected to introduce high levels of absorption in the
Far UltraViolet (FUV). This is qualitatively consistent with the observed rapid rise in extinction toward the
FUV (e.g. Fitzpatrick 2004). Although the properties of the Lyman and Werner transitions have been well studied
in the gas phase of \htwo\ \citep{abgrall1,abgrall2}, we are not aware of any studies of their appearance
in the condensed phase. Similarly, we are not aware of any studies of the properties of \hminus\ in a solid
hydrogen matrix. Quantitative estimates of their contributions must therefore be deferred.

\subsection{Red fluorescence and the ERE}
In \S2.2 we noted that irradiated solid hydrogens are seen to fluoresce in the far red and near infrared.
This fluorescence may be of interest in an astrophysical context as a possible explanation for the Extended
Red Emission (ERE) which has been observed from various types of nebulae and the broader ISM --- see
\citet{wittvijh} for a review of the ERE phenomenon.  With reference to the requirements on the ERE carrier
listed by \citet{wittvijh} we note the following points of similarity between the two phenomena: (i) the spectral
peaks occur in similar places; (ii) the spectra are similarly broad, unstructured and confined to similar wavelength
ranges; (iii) both are photoluminescence phenomena; and (iv) the ERE arises from an
abundant element with Galaxy-wide distribution. There is no doubt that hydrogen is an abundant element with
a Galaxy-wide distribution. We do not yet know if it is found in solid form in astrophysical contexts, but if it is then
it would be a strong candidate for the carrier of the ERE.

\section{Conclusions}
It is remarkable that a single, small hydrogen molecule exhibits most of the strong mid-IR bands which pervade
astronomical environments. This fact alone makes \hd\ of interest to astronomers, even though we cannot yet place
it in our theoretical picture of the interstellar medium with any degree of confidence.  The  7.7 and 8.6$\,{\rm\mu m}$
astronomical bands, which are often the strongest of all the observed UIR features, are absent from our predictions
for \hd. But there are several vibrational modes which we are unable to characterise, leaving open the prospect
that some of those modes are responsible for the bands in question.

Inside a matrix of solid hydrogen, \hd\ is solvated
and the many weak bonds in the resulting cluster are expected to give rise to a large number of FIR vibrational transitions.
Like the mid-IR bands, these cluster modes will be excited following electronic absorption of optical photons,
and we expect a correspondingly close relationship between mid-IR and FIR emissions from the solvated \hd\ molecule.

We are unable to compute the electronic absorption spectrum of solvated \hd\ because it is sensitive to the details of the
solvation, but the floppy nature of the molecule leads us to expect line properties which are broadly consistent with those
observed for the DIBs. And the multitude of weak lines yields an apparent continuum of optical absorption. 

To these phenomena one can add FUV absorption contributed by the solid hydrogen matrix itself and a red fluorescence,
possibly attributable to \hminus. Thus, in an astronomical context, solid hydrogen is expected to manifest itself in a
variety of ways, each of which seems to resemble one of the observed astronomical phenomena collectively
attributed to ``dust.''  We conclude that serious consideration should be given to the possibility that solid \htwo\ is
abundant in the interstellar medium. 

\bigskip 

\acknowledgments
We thank Peter Gill for numerous helpful discussions on the quantum chemical aspects of this work.
MAW thanks Aris Karastergiou and Simon Johnston for feedback on the astrophysical picture,
and AK for introducing MW to VPN. MAW also thanks Sterl Phinney for making available his 1985
unpublished work on the survival of \htwo\ snowballs, Bruce Draine for several useful discussions
on the physics of the ISM, and Dave Anderson for helpful discussions concerning impurities in
\htwo\ ices. Part of this research was undertaken on the NCI National Facility in Canberra,
which is supported by the Australian Commonwealth Government. C.Y. Lin is grateful for support  from
the Australian Research Council's Centres of Excellence programme, via funding allocated to M.L. Coote.

\clearpage

\appendix

\section{Isotopic variants of \hsix}
In \S2 we noted that the dominant isotopomer of \hsix\ in astrophysical environments is usually either \hsix\ or  \hd. In the laboratory, however, conditions may be chosen which favour different isotopomers. Here we tabulate properties of the ten isotopomers which, in addition to \hsix\ and \hd, arise in solids made from mixtures of \htwo\ with HD or ${\rm D_2}$.

In table A1 we list the Zero-Point Vibration Energies of all these isotopomers. These values for the ZPVE include anharmonic corrections, treated as a perturbation according to \citep{zpve}
\begin{eqnarray}
\label{ZPVE}
{\rm ZPVE} & = &
\frac{1}{2}\sum_{i}^{3N-6}\omega_i-\frac{1}{32}\sum_{ijk}^{3N-6}\frac{\phi_{iik}\phi_{kjj}}{\omega_k}
-\frac{1}{48}\sum_{ijk}^{3N-6}\frac{\phi_{ijk}^2}{\omega_i+\omega_j+\omega_k}+\frac{1}{32}\sum_{ij}^{3N-6}\phi_{iijj}+ Z_{kin},
\end{eqnarray}
where the various $\phi$'s are the mode-coupling constants, given by
\begin{equation}
\phi_{ijk} = \frac{\eta_{ijk}}{\sqrt{\omega_i \omega_j \omega_k }},
\end{equation}
and similarly for the quartic constants.
We have not evaluated the three-mode coupling coefficients ($\phi_{ijk}$, see \S3.2) and so we ignore the third term on the right-hand-side of equation (A1).
We also ignore the last term ($Z_{kin}$), associated with rotation, as here we are treating mode \#1 as a vibration on a par with the other modes.

To determine whether or not isotopic condensation occurs it is also necessary to know the ZPVE for \htwo\ and HD or ${\rm D_2}$. From table 39 in the appendix
of \citet{herzberg} we find ZPVE(\htwo)=268.8~meV, ZPVE(HD)=233.7~meV and ZPVE(${\rm D_2}$)=191.4~meV.

In tables A2 and A3 we list the wavenumbers and intensities of the normal vibrational modes of ten isotopomers, to complement the results for \hsix\ and \hd\ given in table 1. Likewise the plots in figures A1 and A2 complement those in figure 3. The theoretical modelling is as described in \S3. We remind readers that the line intensities reported here are computed in the harmonic approximation, even though we have calculated anharmonic corrections for the frequencies of the lines. All transitions in tables A2 and A3 are $v=1\leftrightarrow0$, for the mode in question, and rotation is neglected. The notation for an isotopomer consisting of ${\rm End_1}$, Centre and ${\rm End_2}$ moieties is ${\rm[End_1(Centre)End_2]^+}$.

\begin{deluxetable}{ccccc}
\tabletypesize{\scriptsize}
\tablecaption{Harmonic and anharmonic Zero-Point Vibration Energies for various \hsix\ isotopomers\label{tableA1}}
\tablewidth{0pt}
\tablehead{
\colhead{Isotopomer} & \colhead{} &\colhead{Harmonic ZPVE}  & \colhead{Anharmonic ZPVE} \\
\colhead{} &  \colhead{ } & \colhead{(meV)} &   \colhead{(meV)} \\
}
\startdata
${\rm[H_2(H_2)H_2]^+}$&       &  $\!\!$1010.0   &   $\!\!$1000.9     \\
${\rm[H_2(H_2)HD]^+}$  &       &  964.5    &    955.3    \\
${\rm[H_2(HD)H_2]^+}$  &       &  962.3    &    952.7   \\
${\rm[H_2(HD)HD]^+}$   &       &  916.6     &    906.2    \\
${\rm[H_2(DH)HD]^+}$   &       &  915.9     &    907.5    \\
${\rm[HD(H_2)HD]^+}$   &       &  918.6     &    909.5    \\
${\rm[HD(HD)HD]^+}$     &       &  869.9     &   860.7     \\
${\rm[H_2(H_2)D_2]^+}$&       &  913.6     &   905.8     \\
${\rm[H_2(D_2)H_2]^+}$&       &  912.3    &    903.2    \\
${\rm[H_2(D_2)D_2]^+}$&       &   813.9    &   807.3     \\
${\rm[D_2(H_2)D_2]^+}$&       &   815.9    &   810.2     \\
${\rm[D_2(D_2)D_2]^+}$&       &   714.4    &   709.8     \\
\enddata
\tablecomments{Anharmonic corrections to the ZPVE were computed using equation \ref{ZPVE} with $Z_{kin}$ and $\phi_{ijk}$ terms neglected.}
\end{deluxetable}

\begin{deluxetable}{lrcrrrr}
\tabletypesize{\scriptsize}
\tablecaption{Normal modes of the $D_{2d}$ structure of \hsix\ isotopomers containing H$_2$ and HD groups.\label{tableA2}}
\tablewidth{0pt}
\tablehead{
\colhead{Species} & \colhead{}  & \colhead{Mode} & \colhead{$I_o$} & \colhead{$\omega_o$} & \colhead{$\nu$} \\
\colhead{} & \colhead{} & \colhead{\#} & ${\rm(km\; mol^{-1})}$ & ${\rm(cm^{-1})}$ & ${\rm(cm^{-1})}$\\
}
\startdata
${\rm[H_2(H_2)HD]^+}$\  &   &   12  &   422.7   &   3821.0  &   3530.2  \\
    &   &   11  &   395.6   &   3307.9  &   2973.6  \\
    &   &   10  &   0.6 &   2081.4  &   1774.8  \\
    &   &   ${}\;\;\;\,9^*$   &   8.2 &   1187.6  &   1076.4  \\
    &   &   ${}\;\;\;\,8^*$   &   13.7    &   1169.0  &   1074.3  \\
    &   &   ${}\;\,7$   &   2282.6  &   1004.6  &   1000.3  \\
    &   &   ${}\;\,6$   &   69.2    &   860.7   &   828.7   \\
    &   &  ${}\;\;\;\,5^*$   &   11.4    &   723.7   &   740.8   \\
    &   &   ${}\;\;\;\,4^*$   &   40.3    &   627.3   &   738.4   \\
    &   &  ${}\;\;\;\,3^*$   &   1.5 &   349.0   &   227.5   \\
    &   &   ${}\;\;\;\,2^*$   &   6.2 &   332.6   &   740.8   \\
    &   &   ${}\;\;\;\,1^*$   &   2.2 &   92.9    &   170.6   \\
\\                                      
${\rm[H_2(HD)H_2]^+}$\  &   &   12  &   0.0 &   3858.0  &   3566.2  \\
    &   &   11  &   992.2   &   3777.8  &   3346.5  \\
    &   &   10  &   274.6   &   1842.6  &   1573.2  \\
    &   &  ${}\;\;\;\,9^*$   &   3.6 &   1104.6  &   1024.1  \\
    &   &   ${}\;\;\;\,8^*$   &   19.7    &   1020.9  &   970.0   \\
    &   &   ${}\;\,7$   &   797.1   &   927.3   &   937.5   \\
    &   &   ${}\;\,6$   &   810.5   &   862.1   &   830.0   \\
    &   &   ${}\;\;\;\,5^*$   &   14.9    &   725.7   &   724.6   \\
    &   &   ${}\;\;\;\,4^*$   &   24.8    &   660.7   &   724.6   \\
    &   &   ${}\;\;\;\,3^*$   &   2.9 &   325.0   &   727.5   \\
    &   &   ${}\;\;\;\,2^*$   &   3.0 &   320.2   &   211.2   \\
    &   &   ${}\;\;\;\,1^*$   &   0.0 &   99.2    &   179.2   \\
\\                                      
${\rm[H_2(HD)HD]^+}$\   &   &   12  &   431.3   &   3820.9  &   3541.4  \\
    &   &   11  &   426.1   &   3306.2  &   2970.2  \\
    &   &   10  &   249.8   &   1832.0  &   1624.8  \\
    &   &   ${}\;\;\;\,9^*$   &   2.8 &   1074.7  &   1001.6  \\
    &   &   ${}\;\;\;\,8^*$   &   22.5    &   1018.3  &   925.4   \\
    &   &   ${}\;\,7$   &   1056.9  &   923.7   &   855.7   \\
    &   &   ${}\;\,6$   &   509.2   &   805.4   &   789.5   \\
    &   &   ${}\;\;\;\,5^*$   &   12.4    &   723.5   &   664.3   \\
    &   &   ${}\;\;\;\,4^*$   &   56.1    &   566.8   &   666.8   \\
    &   &   ${}\;\;\;\,3^*$   &   1.0 &   313.4   &   202.9   \\
    &   &   ${}\;\;\;\,2^*$   &   3.2 &   307.5   &   175.4   \\
    &   &   ${}\;\;\;\,1^*$   &   1.8 &   92.8    &   145.8   \\
\\                                      
${\rm[H_2(DH)HD]^+}$\   &   &   12  &   445.4   &   3819.9  &   3558.5  \\
    &   &   11  &   409.7   &   3307.5  &   2979.5  \\
    &   &   10  &   301.1   &   1837.6  &   1611.3  \\
    &   &   ${}\;\;\;\,9^*$   &   4.3 &   1103.3  &   990.7   \\
    &   &   ${}\;\;\;\,8^*$   &   37.6    &   1002.3  &   964.4   \\
    &   &   ${}\;\,7$   &   1446.6  &   866.9   &   926.9   \\
    &   &   ${}\;\,6$   &   29.7    &   857.9   &   778.5   \\
    &   &  ${}\;\;\;\,5^*$   &   21.6    &   658.1   &   694.7   \\
    &   &   ${}\;\;\;\,4^*$   &   64.4    &   615.9   &   694.7   \\
    &   &   ${}\;\;\;\,3^*$   &   0.8 &   314.6   &   675.7   \\
    &   &   ${}\;\;\;\,2^*$   &   3.5 &   298.3   &   270.4   \\
    &   &   ${}\;\;\;\,1^*$   &   1.4 &   92.7    &   163.0   \\
\\                                      
${\rm[HD(H_2)HD]^+}$\   &   &   12  &   6.4 &   3343.6  &   3123.5  \\
    &   &   11  &   677.2   &   3276.9  &   2942.5  \\
    &   &   10  &   0.1 &   2075.5  &   1748.5  \\
    &   &   ${}\;\;\;\,9^*$   &   3.5 &   1167.6  &   1037.9  \\
    &   &   ${}\;\;\;\,8^*$   &   29.3    &   1166.3  &   1041.3  \\
    &   &   ${}\;\,7$   &   2293.7  &   979.4   &   941.7   \\
    &   &   ${}\;\,6$   &   2.5 &   829.3   &   775.8   \\
    &   &   ${}\;\;\;\,5^*$   &   57.8    &   643.0   &   684.1   \\
    &   &   ${}\;\;\;\,4^*$   &   2.5 &   600.6   &   677.0   \\
    &   &   ${}\;\;\;\,3^*$   &   10.7    &   328.3   &   217.3   \\
    &   &   ${}\;\;\;\,2^*$   &   4.0 &   322.4   &   187.6   \\
    &   &   ${}\;\;\;\,1^*$   &   5.6 &   84.9    &   133.5   \\
\enddata
\tablecomments{Wavenumbers are given both with ($\nu$) and without
($\omega_o$) anharmonic corrections. Changes in rotational state of the molecule shift
the lines away from these wavenumbers. Modes marked with an asterisk are strongly
affected by rotation of the end-groups, so these modes
are not good approximations to the eigenmodes of the real molecule (\S3.2.1).}
\end{deluxetable}

\begin{deluxetable}{lrcrrrr}
\tabletypesize{\scriptsize}
\tablecaption{Normal modes of the $D_{2d}$ structure of \hsix\ isotopomers containing H$_2$ and D$_2$ groups.\label{tableA3}}
\tablewidth{0pt}
\tablehead{
\colhead{Species} & \colhead{}  & \colhead{Mode} & \colhead{$I_o$} & \colhead{$\omega_o$} & \colhead{$\nu$} \\
\colhead{} & \colhead{} & \colhead{\#} & ${\rm(km\; mol^{-1})}$ & ${\rm(cm^{-1})}$ & ${\rm(cm^{-1})}$\\
}
\startdata
${\rm[H_2(H_2)D_2]^+}$\ &   &   12  &   465.5   &   3819.2  &   3476.3  \\
    &   &   11  &   205.2   &   2703.8  &   2474.5  \\
    &   &   10  &   2.6 &   2075.0  &   1785.0  \\
    &   &   ${}\;\;\;\,9^*$   &   9.5 &   1185.8  &   1064.7  \\
    &   &   ${}\;\;\;\,8^*$   &   0.9 &   1153.5  &   1052.5  \\
    &   &   ${}\;\,7$   &   2126.6  &   996.9   &   979.4   \\
    &   &   ${}\;\,6$   &   250.3   &   760.0   &   749.0   \\
    &   &   ${}\;\;\;\,5^*$   &   9.5 &   721.4   &   708.7   \\
    &   &   ${}\;\;\;\,4^*$   &   2.2 &   581.4   &   708.7   \\
    &   &   ${}\;\;\;\,3^*$   &   4.2 &   341.2   &   699.7   \\
    &   &   ${}\;\;\;\,2^*$   &   2.2 &   310.5   &   248.2   \\
    &   &   ${}\;\;\;\,1^*$   &   0.0 &   88.2    &   149.4   \\
\\                                      
${\rm[H_2(D_2)H_2]^+}$\ &   &   12  &   0.0    &	  	3857.8   &	  	3483.6  \\
    &   &   11                    &   1039.1   &	  	3776.2   &	  	3350.8  \\
    &   &   10                    &  0.0    &	  	1521.4   &	  	1361.6  \\
    &   &   ${}\;\;\;\,9^*$   &   17.6   &	  	 923.5   &	  	874.8   \\
    &   &   ${}\;\;\;\,8^*$   & 17.6   &	  	923.4   &	  	874.8  \\
    &   &   ${}\;\,7$           &  1331.9   &	  	832.0   &	  	773.9  \\
    &   &   ${}\;\,6$           &  0.0   &	  	         880.3   &	  	880.7   \\
    &   &   ${}\;\;\;\,5^*$   &  16.1   &	  	648.1   &	  	665.0   \\
    &   &   ${}\;\;\;\,4^*$   &  16.1   &	  	648.0   &	  	664.6   \\
    &   &   ${}\;\;\;\,3^*$   &   9.4   &	  	303.4   &	  	182.9   \\
    &   &   ${}\;\;\;\,2^*$   &  9.4   &	  	303.4   &	  	184.0  \\
    &   &   ${}\;\;\;\,1^*$   &  0.0   &	  99.2   &	  	154.4   \\
\\                                      
${\rm[H_2(D_2)D_2]^+}$\ &   &   12  &  504.5	   &  3817.7   &  	3506.9    \\
    &   &   11                   &    252.9   &  	2700.8	   &  2470.3  \\
    &   &   10                   &   6.1   &  	1497.8	   &  1345.7  \\
    &   &   ${}\;\;\;\,9^*$   &  21.7	   &  920.8   &  	850.3 \\
    &   &   ${}\;\;\;\,8^*$   & 1.5   &  	846.7   &  	826.1  \\
    &   &   ${}\;\,7$           &  682.0   &  	852.2   &  	804.4 \\
    &   &   ${}\;\,6$           & 587.7	   &  690.0   &  	660.2  \\
    &   &   ${}\;\;\;\,5^*$   &  9.8   &  	640.8   &  	610.9  \\
    &   &   ${}\;\;\;\,4^*$   &  12.1   &  	520.3   &  	583.7 \\
    &   &   ${}\;\;\;\,3^*$   &  1.9   &  	283.5   &  	240.9   \\
    &   &   ${}\;\;\;\,2^*$   &  2.0 &	270.7   &  	164.0   \\
    &   &   ${}\;\;\;\,1^*$   &  0.0 &	88.2   &  	122.9     \\
\\                                      
${\rm[D_2(H_2)D_2]^+}$\ &   &12  &   0.0 &   2730.1  &   2600.0  \\
    &   &   11  &   390.6   &   2677.7  &   2443.0  \\
    &   &   10  &   0.0 &   2063.6  &   1718.6  \\
    &   &   ${}\;\;\;\,9^*$   &   2.1 &   1149.0  &   1015.2  \\
    &   &   ${}\;\;\;\,8^*$   &   2.1 &   1149.0  &   1015.2  \\
    &   &   ${}\;\,7$   &   2321.7  &   927.9   &   876.9   \\
    &   &   ${}\;\,6$   &   0.0 &   648.7   &   629.2   \\
    &   &   ${}\;\;\;\,5^*$   &   0.5 &   574.9   &   626.2   \\
    &   &   ${}\;\;\;\,4^*$   &   0.5 &   574.8   &   625.9   \\
    &   &   ${}\;\;\;\,3^*$   &   9.6 &   297.5   &   223.9   \\
    &   &   ${}\;\;\;\,2^*$   &   9.6 &   297.4   &   224.5   \\
    &   &   ${}\;\;\;\,1^*$   &   0.0 &   70.1    &   98.5    \\
\\                                      
${\rm[D_2(D_2)D_2]^+}$\ &   &   12  &       0.0 & 	2729.2  &		2540.4    \\
    &   &   11                    &  477.0 &	2673.3 &		2451.7    \\
    &   &   10                    &   0.0 &		1476.6 &		1311.5 \\
    &   &   ${}\;\;\;\,9^*$   &   3.5 &		841.4 &		773.7  \\
    &   &   ${}\;\;\;\,8^*$   &   3.5	 &	841.4 &		773.7  \\
    &   &   ${}\;\,7$           &  1210.3 &	720.2 &		687.2  \\
    &   &   ${}\;\,6$           &   0.0 &		641.6 &		605.3   \\
    &   &   ${}\;\;\;\,5^*$   &  6.8 &		513.6 &		537.9   \\
    &   &   ${}\;\;\;\,4^*$   &  6.8 &		513.6 &		537.7  \\
    &   &   ${}\;\;\;\,3^*$   &  0.1 &		252.0 &		213.6  \\
    &   &   ${}\;\;\;\,2^*$   &  0.1 &		252.0 &		214.0  \\
    &   &   ${}\;\;\;\,1^*$   &  0.0  &	         70.1 &		85.6 \\
\enddata
\tablecomments{Wavenumbers are given both with ($\nu$) and without
($\omega_o$) anharmonic corrections. Changes in rotational state of the molecule shift
the lines away from these wavenumbers. Modes marked with an asterisk are strongly
affected by rotation of the end-groups, so these modes
are not good approximations to the eigenmodes of the real molecule (\S3.2.1).}
\end{deluxetable}

\begin{figure}
\figscaletwo
\plotone{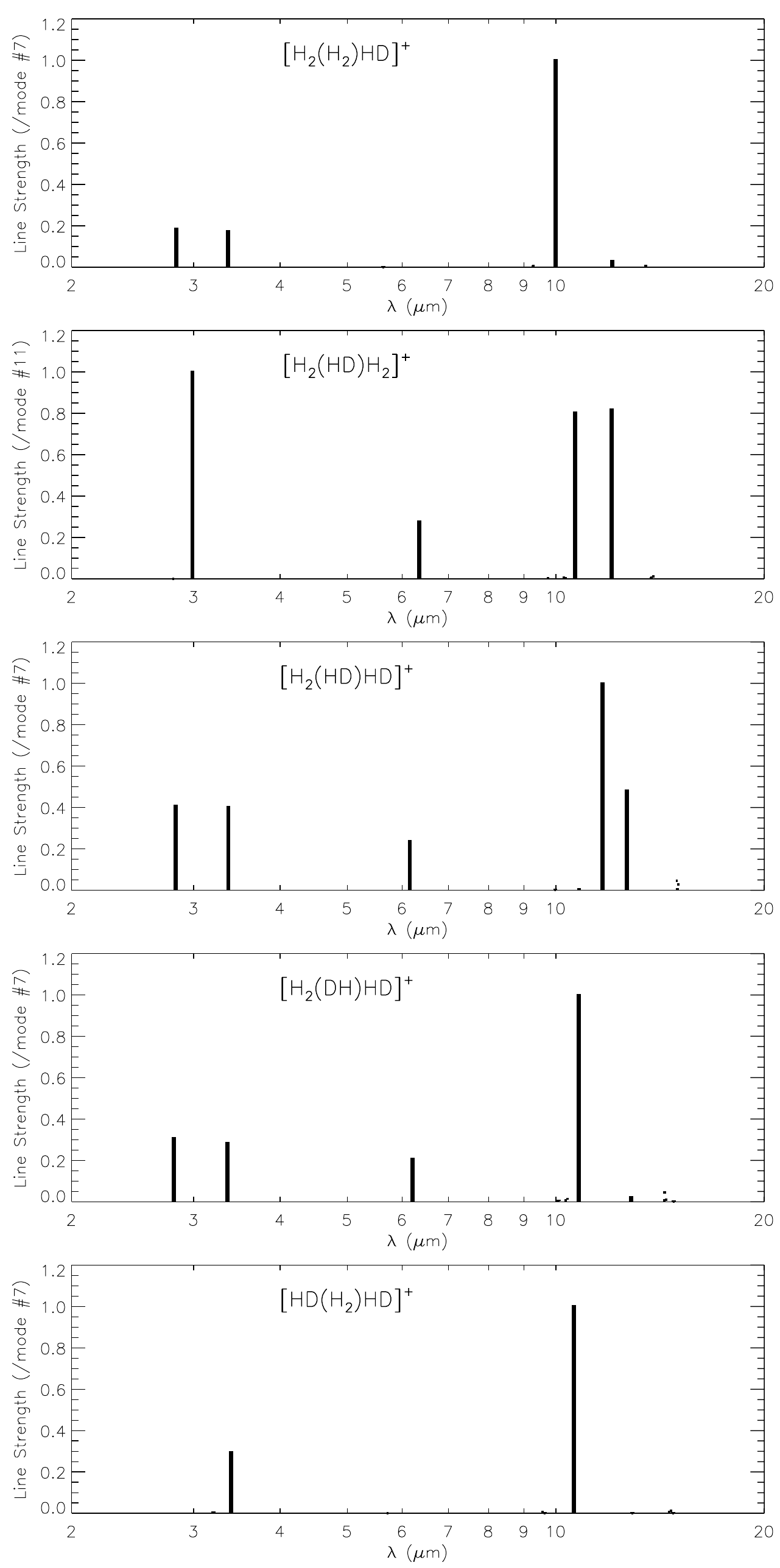}
\caption{Stick diagram of the mid-IR lines of five isotopomers of \hsix. The line strengths are normalised
to that of mode \#7 or \#11 for each molecule. The E-type modes of the $D_{2d}$ structure are not good approximations to the
actual eigenmodes of the molecules (see \S3.2.1) and they are therefore plotted with dotted lines.}
\end{figure}

\begin{figure}
\figscaletwo
\plotone{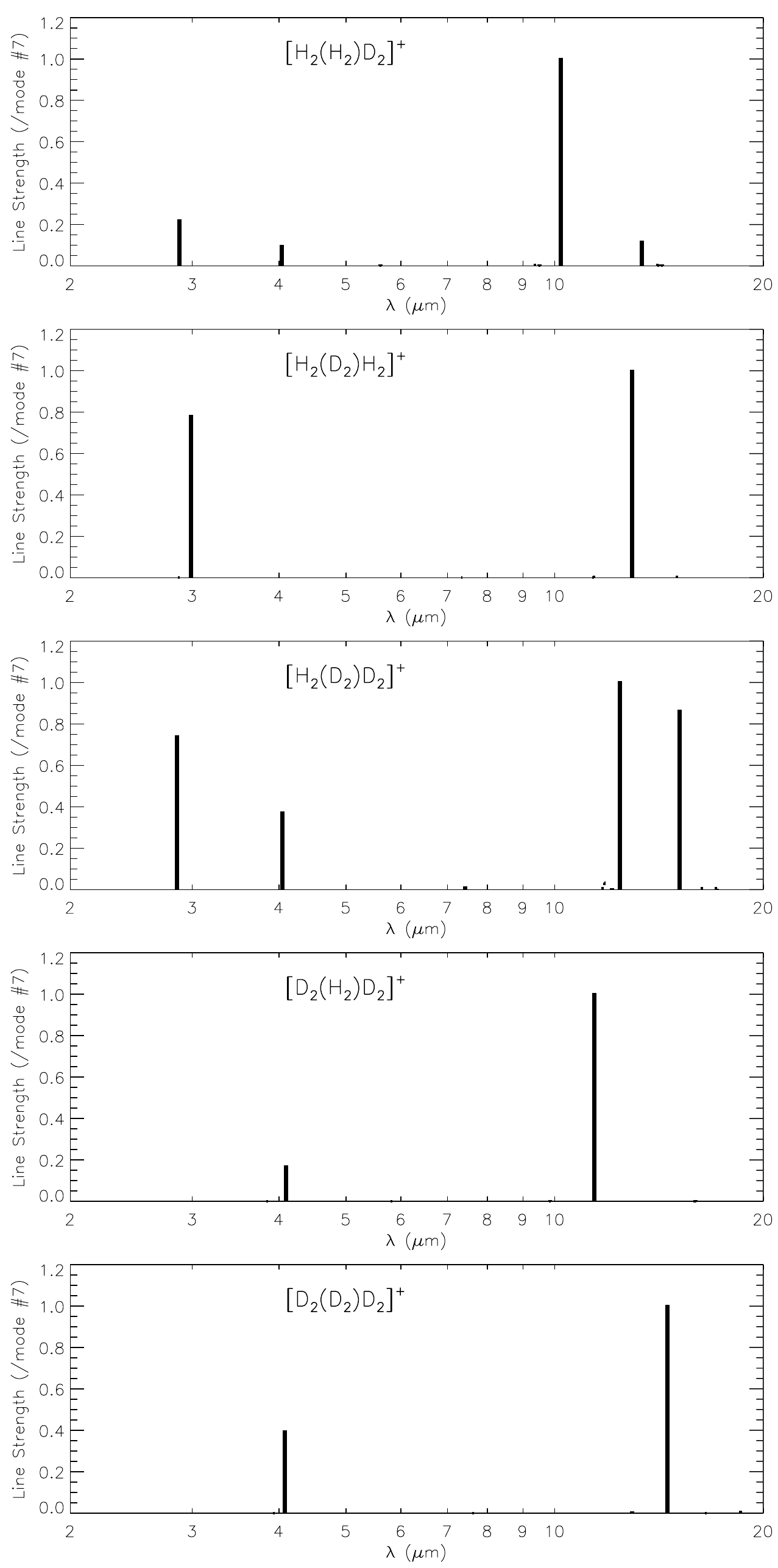}
\caption{Stick diagram of the mid-IR lines of five isotopomers of \hsix. The line strengths are normalised
to that of mode \#7 for each molecule. The E-type modes of the $D_{2d}$ structure are not good approximations to the
actual eigenmodes of the molecules (see \S3.2.1) and they are therefore plotted with dotted lines.}
\end{figure}

\end{document}